# Quelques réflexions sur la conception du monde de Erwin Schrödinger

Mário J. Pinheiro
Departamento de Física e Centro de Física de Plasmas, Instituto Superior Técnico,
Av. Rovisco Pais, 1049-001 Lisboa, Portugal

Abstract: Some personal reflections are made about the philosophical posture of Erwin Schrödinger, as exposed in one of his book "Ma Conception du Monde". It is argued that the Taoist view of the world, as pursued by Schrödinger, although characterized by the dialectical struggle between opposite truths does not contain the temporal dimension, the unique which allows an evolving understanding of phenomena and objects that compose this world in constant evolution. In our point of view, only a genetic view - which is missing in the actual scientific paradigm - will bring a more deep explanation and understanding of Reality.

1. Introduction

On voudrait dénuer le sens de la pensée d'un physicien de la taille de Erwin Schödinger. Par la place qu'il occupe dans l'histoire de la Physique, par sa contribution à une vision de la Nature tout à fait nouvelle où les fondements de la microphysique se sont bauchés, et par une position religieuse, mystique, coïncidente à maints égards avec d'autresphysiciens contemporains, Schrödinger suscite notre curiosité.

Dans la préface de son livre « Ma conception du Monde » (Schrödinger, 1982) on voit à quel point sa recherche des choses prend une allure sérieuse. Il y est dit du poids qui ont joué les lectures de Spinoza, Schopenhauer, Mach, Richard Semon et Richard Avenarius. D' ailleurs, son livre a été écrit en deux périodes bien lointains l'un de l'autre : « La Quête du Chemin » a été écrit en automne 1925 et «Qu'est ce qui est réel ? » en 1960.

Schrödinger, avec De Broglie, Einstein, Planck ont toujours critiqué dans la Mécanique Quantique le caractère non définitif. Cette école de pensée a été considérée matérialiste pour autant, et en opposition à des physiciens comme Heisenberg, Bohr, Jordan, Pauli qui defendèrent le point de vue que la Mécanique Quantique révélait l'a-causalité des événements et la raison humaine touchait ses limites. On reprochait à ceux-ci de régresser dans l'idéalisme. Les deux écoles opposaient des perspectives si divergentes que tout débat a été presque impossible et les tenants « idéalistes » sont arrivés à emporter l'acceptation générale de son interprétation de la Mécanique Quantique. Le vainquer du débat a été le «der Compenhagen Geist », selon une expression d'Heisenberg.

Une question qu'on se pose c'est la suivante : pourquoi Schrödinger cherche une vision du monde si loin de nous ? Question licite, puisqu'en Occident il y a eu des penseurs remarquables qui nous ont laissé des conceptions dialectiques de la Nature, comme Karl Marx, L. Feuerbach, Nietzsche.

2. L'Orientation philosophique de Schrödinger

Dans la première partie du livre, intitulé « La Quête du Chemin », on trouve au Ch. V, « La Vision Fondamental du Vedânta » où Schrödinger révèle comme il voit personnellement l'issue du problème posé avant et qui étaient : Le Moi, Le Monde, la mort et la Pluralité.

Les Veda (sagesse) se trouvent au sommet de la littérature religieuse Hindoue et sont une collection de quatre livres, tous écrits en sanscrit : Rgveda (hymne aux dieux) ; Yajurveda (formule des sacrifices) ; samaveda (cantiques sacrificiel) ; Atharvaveda (formules magiques). D'autres écrits les rejoignent, comme les nombreux Brâhmana (textes sacrificiels) et les traités philosophiques ésotériques. Les textes sacrés des Veda sont généralement considérés comme une révélation directe de Brahman.

Bien sûr, on cherche ici à distinguer son chemin personnelle de sa méthode de pensée et à répondre à des questions tel que: a) Qu'est-ce qu'il cherche à connaître ; b) quelle est son attitude intellectuel - ce qu'il accepte, ce qu'il refuse.

On peut voir qu'il y a une espèce de identification : le moi -> être unique, la mort -> immortalité, pluralité -> il n'accepte pas les différences. Il a une pensée de continuité. Ce qu'il y a derrière sa pensée, dans son inconscient, c'est une attitude de syncrétisme. Il n'y a pas de rupture, c'est une pensée de continuité, ce n'est pas une pensée dialectique. C'est ça que dirige son chemin, sa recherche de sensorimoteur, une recherche de ressemblances. Mais pour connaître et pour expliquer, il faut voir les différences et les ressemblances entre les phénomènes, c'est une pensée que je partage avec Wallon. La philosophie orientale résout alors très bien ses problèmes intellectuels, vu qu'il ne cherche pas une explication dialectique du monde. Le caractère de tous les philosophes mystiques c'est de voir une cause unique en tout.

Il y a une ressemblance de points de vues entre la microphysique et la psychologie, selon C. G. Jung. On sait que l'idée fondamentale de Jung est celle d'inconscient collectif et c'est en ceci que sa théorie diffère de celle de Freud, où l'inconscient est individuel. Or, selon Jung, l'inconscient est le siège de forces puissantes que laissent son empreinte indélébile dans les domaines si divers de l'activité humaine, comme la religion et la mythologie, l'art et la physique, comme a ressorti des discussions entre Jung et Wolfgang Pauli.

Le concept de complémentarité, introduit par Niels Bohr dans la Physique, semble avoir son analogue dans la psychologie, d'après Jung, dans le couple d'opposées conscient-inconscient.

L'inconscient collectif semble donc, dans la conception jungienne, être comme qu'un immense réservoir où l'humanité va puiser ses idées, qui peuvent être créatives et amener à son développement, ou destructrices et barrer l'évolution de l'esprit (conscient et inconscient, puisque les deux semblent être différents).

Il agit comme une espèce de mémoire supra individuelle, de telle sorte agissant que «…aucun Moi n'est isolé.» et «…le Moi n'est pas tant *lié* à ce qui est arrivé à ses ancêtres, il n'est pas tant le produit […] que plutôt, au sens le plus strict, <u>la même chose que tout cela</u> : il est son prolongement absolu et immédiat, comme le moi à l'âge de cinquante ans est le prolongement de celui de quarante ans» (p.53 de l'oeuvre cité).

Mais il nous semble que le problème ne devrait pas être si généralisée ; il y a pour autant une nuance entre un homme et un végétal et ne serait-il une généralisation un peu forcé de voir tout identique, tout une seule chose?

Le principe de la complémentarité exige qu'il y ait une étendue du champ de connaissance. Il faut étudier un événement physique d'une façon dynamique et donc il faut une méthode pluridimensionnelle pour retenir l'aspect dynamique d'un phénomène physique. Par pluridimensionnelle on veut dire différents champs de connaissance. Et ceci ce n'est pas la pensée de la science physique et ceci est étonnant vu que Schrödinger est un physicien.

Connaître c'est voir les différents aspects d'un phénomène (décrire tout ce qu'il y a des differents, des dissemblances). Pour connaître il faut expliquer, connaître son origine, son développement, sa fin - connaître ses transformations.

À la limite on pourrait dire que tout est une seule chose : énergie, Dieu, Nature,…Mais on perdrait la capacité de différencier, de séparer, de catégoriser, ce que serait déjà le prélude d'une intelligence en éveil. D'ailleurs, le contexte n'esp pas le même, quand Schrödinger donne l'exemple du prolongement existant entre un homme de quarante et cinquante ans ; c'est bien plus commun et naturel, tel continuité, mais déjà il existe une rupture abyssale entre un enfant de cinq ans et un adulte de cinquante ans.

Au Ch. VI, «Introduction ésotérique à la pensée scientifique», Schrödinger suive donc la thèse d'une mémoire supra individuelle, responsable d'une «echphorie» (une sorte d'atavisme) de très anciens engrammes. Aucun moi n'est isolé. Et si dans la philosophie occidentale on concevait que la mort n'était pas nécessairement la fin de l'essence de la vie, qui pourrait continuer dans un autre monde supérieur, alors on s'étonne qu'elle n'ait pas conçu que la naissance d'un être pouvait être multiple.

Dans l'Hindouisme, le Cosmos est éternel et en perpétuelle évolution tandis que les innombrables mondes qui le composent, les œufs de Brahmâ (Brahmanda), sont assujetties au cycle périodique de la naissance, vie et mort. Tous les êtres vivants ont une âme immortelle mais sont enveloppés par la matière et obligés à s'abandonner à ce cycle incessant, par la loi des retributions, qui juge d'après les bonnes ou mauvaises actions. Ainsi, finalement, partagant cette conception hindouiste, ce n'est qu'en apparence (maya) que le monde semble multiple, puisqu'en réalité, un être capable d'atteindre un niveau supérieur de contemplation, arrive à travers ce voile obscur, l'unité de toutes les choses. Cette idée est à maintes reprises évoquées par Schrödinger et c'est dans le ch. IV qu'on peut cerner le vrai problème, celui de la pluralité et de l'unité. C'est ici que l'on trouve les questions posées.

Schrödinger attribue l'origine d'une conception dualiste de la relation esprit et matière dans le langage et l'éducation. Cette partition serait une source de difficultés pour notre compréhension du monde, puisque sous le point de vue de l'Esprit, une détermination causale des lois qui régissent les phénomènes naturels, matériels, doit porter des fautes empêchant une connaissance objective des choses matérielles ; de l'autre coté, il nous échapperait une compréhension des influences causales des choses de l'esprit. Clairement, Schrödinger n'esp pas idéaliste. Son attitude c'est justement de rejeter la connaissance de l'Essence. Il ne résiste pas aux différences. Mais il faut bien voir que l'univers se crée en donnant des transformations des choses. C'est un défaut de méthode de pensée en Schrödinger.

Il n'y aurait de solution convenable qu'en renonçant à ce dualisme. Mais est-ce le dualisme une exigence de dépassement du niveau de compréhension ? Vouloir renoncer au dualisme simplement pour éviter une dualité de l'être et la pensée, la matière et l'esprit, n'est il pas le reflet d'une pensée statique, immobilisée, justement parce que la dualité traduit deux niveaux différents de compréhension ? On trouve qu'en Schrödinger il n'y a as pas de dimension temporelle. On dirait qu'il y a une attitude de stabilisation ; on cherche à stabiliser un objet, à l'immobiliser, pour meilleure sécurité de l'observateur, et le meilleur c'est de le voir comme un tout unique.

Pour une perspective évolutive, il faudrait voir l'objet dans ses transformations. Mais de toute façon, une attitude de stabilisation est une attitude scientifique, en particulier c'est la situation la plus convenable pour une expérimentation scientifique, quand on répète le même processus afin de vérifier la constance de certains paramètres en vue d'autres.

Les enfants eux aussi, répètent les mêmes mouvements, frappant un jouet contre quelque chose pour voir comment il est fait. L'enfant nécessite répéter et observer que rien ne s'altère et c'est en ce moment que la connaissance est acquise par stabilisation et pour la sécurité de l'enfant.

Dans la deuxième partie Ch.II, intitulé «La compréhension verbale et notre perception commune du monde on voit à quel point le langage est insuffisant pour une compréhension du Monde. Seulement, par lui c'est sûr qu'on n'a pas le droit d'espérer tout comprendre, tout expliquer. Le langage est limité. Mais ce n'est pas uniquement par lui que le réel peut être compris.
 Une perception commune du monde, ne serait-elle la fin de l'Art ? Si tout le monde voyait exactement les mêmes choses, l'artiste serait un personnage inutile, tout création impossible. N'est-ce pas la différence, les paradoxes, les labyrinthes, l'illogisme des choses que propulserait aussi l'humanité en progrès ?
C'est si important d'établir les similitudes de deux objets que leurs dissemblances, que leurs différences, afin qu'un enfant commence à comprendre ces objets.
Il me semble impossible d'atteindre une compréhension du réel, dans une seule perspective syncrétique. La conscience jaillit quand on voit les différences. Dans la confusion inextricable des donnés, des faits, une conscience ne peut pas s'affirmer.
La psychologie a prouvée l'existence de divers niveaux de conscience : conscience corporel, des états internes, de mouvement, de l'autre,…jusqu'à l'infini. Il ne faudrait pas oublier ceci. Essayer d'expliquer l'homme en se rapprochant à des modèles physiques pose des sérieuses difficultés ; l'homme est irréductible à un chat.
Quand l'homme ne comprends pas, est-ce qu'il tend à croire ?
Dans ce brouillard diffus, indéfini, l'homme est au niveau affectif et il se peut bien que le langage soit le reflet de cette émotion au début de sa formation. De cet effort à la recherche d'une compréhension du réel, de ce qui nous entoure et existe indépendamment de nous en une sorte de cohérence intrinsèque, le langage devient-il alors descriptif et conceptuel ?

A propos de croyance, les observations de Lawrence Kohlberg's (Kohlberg, 1987) chez les enfants d'un groupe aborigène malaisien, en Formosa, constatent la présence de la notion de conservation de la masse, poids, à l'âge de 7-8 ans, mais que cette notion est partiellement perdue entre 11 et 15 ans. Selon le chercheur: «*…the loss did not seem to be a genuine regression but an uncertainty about thrusting their own judgement, that is, there was an increase in "don't know" response. Apparently, adolescent confrontation with adult magical beliefs led them to be uncertain of their natural physical beliefs, whether or not they were in direct conflict with the adult ideology*».
Qu'est-ce qu est réel ? Cet effort en vue d'une réponse reste dans une perspective imprécise dans le texte de Erwin Schrödinger. Traditionnellement nous pourrions opposer les deux points de vue, celui de l'empirisme conforme on donne une prépondérance aux actions extérieurs, et celui du rationalisme, quand de l'activité mentale on part à la connaissance du monde. L'opposition intrinsèque à ces deux positions philosophiques difficulte son rapprochement, impliquant des graves conséquences pour l'activité scientifique. En plus, une répulsion mutuelle, amène les empiristes vers une doctrine matérialiste et les rationalistes suivent les voies de l'Idéalisme. Nous avons alors une substitution des théories sur la nature des connaissances par des théories sur la nature de l'être. Et ici, la tentation d'un réductionnisme est peut-être très forte, la nature de l'être devant être unique, et ainsi le matérialisme réduit cette nature aux forces physiques et l'idéalisme aux constructions de l'esprit.

Le criticisme et le positivisme on envisagé, chacun à sa manière, la solution de cette opposition doctrinale en séparant des discussions le domaine des faits. Ils reconnaissent que par les simples faits, on reste loin des fondements et substance de l'être en soi. Mais, néanmoins, on pourrait établir des relations invariables qui pourrait nous prédire les événements, permettre son contrôle et sa production.

Rappelons nous que la formulation mathématique de la théorie quantique est basée sur l'exigence de la Relativité Restreinte, que toutes les lois de la Physique doivent respecter les transformations covariantes de Lorentz.

Mais à quel point la science est-elle positiviste ? Il faudrait le mieux le définir, car l'avènement de la Théorie Quantique, dont Erwin Schrödinger a été l'un des fondateurs, a introduit une nouvelle conception où l'homme, observateur physique, n'est plus complètement indépendant des événements qu'il observe, puisque object-observateur sont liés de telle sorte qu'une incertitude est introduite dans nos mesures.

Cette idée conduit l'auteur à croire à une perspective globale des choses, toutes les choses n'étant que le reflet multiple d'un être unique. Et il découle une Éthique de remplacement de ce monde qui nous est commun et dans lequel les limites de chacun ne sont pas précisées, de sa nature même, reflet d'une Intelligence Métaphysique et Mystique. N'a pas dit Einstein «le champ est l'unique réalité ?».

Cette intelligence Métaphysique remplacerait une intelligence formée par la discipline scientifique, puisque la science n'est pas capable d'accéder l'homme à une explication et compréhension du Réel.

Ces limites qu'on découvre dans la science, viennent induire un phénomène de rapprochement entre la culture occidentale et l'Orient mystique. En particulier, les physiciens contemporains retrouvent dans l'Hindouisme et le Taoïsme, deux sources enrichissantes pour une autre vision du réel, en particulier la microphysique, qu'est en dehors d'une perception sensorielle de l'organe des sens.

Selon un de ses représentants, Fritjof Capra, «la finalité principale de la mystique orientale est d'expérimenter tous les phénomènes du monde comme des manifestations de la réalité ultime» (Capra, 1981). Le Taoïsme conçoit déjà la réalité comme intrinsèquement dynamique. Sa perspective n'est plus statique et l'harmonie universelle dérive d'un flux continue d'énergies du Yin et Yang.

Le fondateur du Taoïsme, Lao-tseu (ce qui signifie maître ancien) a crée une doctrine basée sur une mystique individuelle et passive. Elle s'exprime théoriquement en des conceptions cosmogoniques amenant à une philosophie de la Nature, dont les analogie avec la Mécanique Quantique semblent frappantes, selon certains physiciens. En pratique, ses disciples suivent une voie d'indifférence aux passions et la méditation de l'origine des choses. Le Taoïsme propose au sage qu'il approfondisse sa connaissance, mais d'une façon spontanée, innocente, sans passions. Son action sur la Nature doit être spontanée, suivre la Nature même, et non s'opposer à elle, d'une forme active et utilitaire. Cet état d'innocence c'est le Tao même. Pour l'anecdote, les légendes relatent que Lao-tseu a écrit l'œuvre la plus importante Tao Te King (« De la loi Universelle et ses Effets ») par demande d'un commandant d'un poste de frontière quand il voulait quitter la Chine assis sur un bœuf noir, prévoyant la chute de l'empire.

### 3. Épilogue

Il semble, donc, que la science abandonne avec sa tradition positiviste. Néanmoins, la valeur des nouvelles perspectives qui s'ouvrent ainsi à la science positiviste restent à mon avis douteuses, puisque sa capacité explicative restent pratiquement nulle.
Quoique le Taoïsme présente une vision intéressante des Opposés, il ne comprend pas les dimension temporelles, l'unique que possibiliterait une compréhension évolutive des phénomènes, des choses qui composent ce monde et que sont loin d'être statiques. Une vision génétique, il me semble, fait défaut à la science, et ce ne peut qu'être une grave lacune dans notre explication et compréhension du Réel.
D'ailleurs, on pense que par la seule discipline scientifique, on ne pourra pas mieux saisir les fondements de la substance réelle. Comme Henri Wallon (Wallon, 1970) a si bien exprimé: « La différence intellectuel entre l'age du totem et celui de la science est moins de niveau que de matériel et de technique idéologiques. Les puissance invisibles du primitif sont évidemment sans ressemblances avec les forces que mesure le physicien, mais à leur façon elle jouent le même rôle ».

Jean Granier (Granier, 1966), dans une thèse profonde sur «Le Problème de la Vérité dans la Philosophie de Nietzche» écrit sur «cette manie pour la logique, cette conviction selon laquelle la vertu est, par essence, démontrable, cette apologie de la dialectique comme seule méthode de réflexion, refléctent donc seulement l'idiosyncrasie des faibles, chez qui la spontanéité créatrice est à ce point étiolée qu'ils ont constamment besoin des béquilles de la logique pour se mouvoir dans l'existence». Dans le «Crépuscule des idoles», Nietzche dit : «ce qui a besoin d'être démontré pour être cru, ne vaut pas grand chose».

Connaître le Réel ne peut être uniquement l'œuvre solitaire de la science, mais doit être surtout le travail incessant et infatigable du mental.